\documentclass{aastex}

 \newcommand{\RXTE}{\emph{Rossi X-ray Timing Explorer}}
 \newcommand{\rxte}{\emph{RXTE}}
 
 \newcommand{\asm}{ASM}
 \newcommand{\PCA}{Proportional Counter Array}
 \newcommand{\pca}{PCA}
 
 \newcommand{\PCUS}{Proportional Counter Units}
 
 \newcommand{\pcus}{PCUs}
 \newcommand{\HEXTE}{High Energy X-ray Timing Experiment}
 \newcommand{\hexte}{HEXTE}

 \newcommand{\sax}{\emph{Beppo}SAX}
 \newcommand{\ginga}{\emph{Ginga}}
 \newcommand{\exosat}{\emph{EXOSAT}}
 \newcommand{\rosat}{\emph{ROSAT}}
 \newcommand{\uhuru}{\emph{UHURU}}
 
 \newcommand{\tenma}{\emph{Tenma}}

 \newcommand{\copernicus}{\emph{Copernicus}}
 
 \newcommand{\bbxrt}{\emph{BBXRT}}

 \newcommand{\CRSF}{Cyclotron Resonance Scattering Feature}
 \newcommand{\CRSFs}{Cyclotron Resonance Scattering Features}
 
 \newcommand{\crsf}{CRSF}
 \newcommand{\crsfs}{CRSFs}
 \newcommand{\flc}{FLC}
 \newcommand{\NPEX}{Negative and Positive power-law Exponential}
 \newcommand{\npex}{NPEX}
 
 \newcommand{\fdco}{FDCO}
 
 \newcommand{\plcut}{PLCUT}
 \newcommand{\xray}{X-ray}
 
 \newcommand{\fasebin}{\emph{fasebin}}

 \newcommand{\cmsq}{\ensuremath{\rm{cm^2}}}
 \newcommand{\pcmsq}{\ensuremath{\rm{cm^{-2}}}}
 \newcommand{\flux}{\ensuremath{\rm{ergs\,cm^{-2}\,s^{-1}}}}
 \newcommand{\lum}{\ensuremath{\rm{ergs\,s^{-1}}}}
 \newcommand{\mom}{\ensuremath{\rm{G\,cm^{3}}}}
 \newcommand{\cps}{\ensuremath{\rm{cts\,s^{-1}}}}
 \newcommand{\cpspcu}{\ensuremath{\rm{cts\,s^{-1}\,PCU^{-1}}}}
 \newcommand{\wsim}{\ensuremath{\sim}}
 \newcommand{\wpm}{\ensuremath{\pm}}
 \newcommand{\wchi}{\ensuremath{\chi^{2}}}
 \newcommand{\wchired}{\ensuremath{\chi^{2}_{\rm{red}}}}

 \newcommand{\wgamma}{\ensuremath{\Gamma}}
 \newcommand{\nh}{\ensuremath{\mathrm{N}_{\mathrm{H}}}}
 \newcommand{\ecut}{\ensuremath{E_{\rm{cut}}}}
 \newcommand{\efold}{\ensuremath{E_{\rm{fold}}}}
 \newcommand{\ecy}[1]{\ensuremath{E_{\rm{c#1}}}}
 \newcommand{\wcy}[1]{\ensuremath{\sigma_{\rm{c#1}}}}
 \newcommand{\dcy}[1]{\ensuremath{D_{\rm{c#1}}}}

 \newcommand{\msol}{\ensuremath{M_{\odot}}}
 \newcommand{\degree}{\hbox{\ensuremath{^\circ}}}

 \newcommand{\ex}[1]{\ensuremath{10^{#1}}}
 \newcommand{\dx}[1]{\ensuremath{\times10^{#1}}}

 \newcommand{\plcuteq}{\ensuremath{
   \mathrm{PLCUT}(E) = A\ E^{-\Gamma}\times\
   \cases{1&$(E\leq\ecut)$ \cr {\rm e}^{-(E-\ecut)/\efold}&$(E>\ecut)$}
 }}

 \newcommand{\crsfeq}[1]{\ensuremath{
    \ecy{#1} = 11.6 B_{12} (1 + z)^{-1}
 }}

 \newcommand{\gabseq}[1]{\ensuremath{
    \tau(E) = \dcy{#1}{\rm e}^{(E-\ecy{#1})^{2}/(2\wcy{#1}^{2})}
 }}

 \newcommand{\widtheqfwhm}{\ensuremath{
    \Delta E_{\rm FWHM} \approx \ecy{}
    \left(\frac{8\,\ln(2)\,kT_{\rm e}}{m_{\rm e}c^{2}}\right)^{1/2}
    \left|\cos(\theta)\right|
 }}

\begin{document}


\title{Discovery of a Cyclotron Resonant Scattering Feature in
the \emph{RXTE} Spectrum of 4U~0352+309 (X~Per)}

\author{W. Coburn\altaffilmark{1},
 W. A. Heindl\altaffilmark{1},
 D. E. Gruber\altaffilmark{1},
 R. E. Rothschild\altaffilmark{1},
 R. Staubert\altaffilmark{2},
 J. Wilms\altaffilmark{2},
 I. Kreykenbohm\altaffilmark{2}
}

\altaffiltext{1}{Center for Astrophysics and Space Sciences, Code
0424, University of California at San Diego, La Jolla, CA,
92093-0424, USA}

\altaffiltext{2}{Institut f\"{u}r Astronomie und Astrophysik,
Astronomie, University of T\"{u}bingen, Waldh\"{a}user Strasse 64,
D-72076 T\"{u}bingen, Germany}


\begin{abstract}

We have discovered a \wsim29\,keV \CRSF\ (\crsf) in the \xray\
spectrum of 4U~0352+309 (X~Per) using observations taken with the
\RXTE. 4U~0352+309 is a persistent low luminosity ($L_{X} =
4.2\times10^{34}$\,\lum) X-ray pulsar, with a 837\,s period and
which accretes material from the Be star X~Per. The X-Ray
spectrum, unusual when compared to brighter accreting pulsars,
may be due to the low mass accretion rate and could be typical of
the new class of persistent low luminosity Be/X-Ray binary
pulsars. We attempted spectral fits with continuum models used
historically for 4U~0352+309, and found that all were improved by
the addition of a \crsf\ at $\sim29$\,keV. The model that best fit
the observations is a combination of a $1.45\pm0.02$\,keV
blackbody with a $5.4\times10^{8}\,\rm{cm}^{2}$ area, and a
power-law with a $1.83\pm0.03$ photon index modified by the
\crsf. In these fits the \crsf\ energy is
$28.6_{-1.7}^{+1.5}$\,keV, implying a magnetic field strength of
$2.5(1+z)\times10^{12}$\,G in the scattering region (where $z$ is
the gravitational redshift). Phase resolved analysis shows that
the blackbody and cyclotron line energies are consistent with
being constant through the pulse.

\end{abstract}



\section{Introduction}\label{sec:intro}

The \xray\ source 4U~0352+309 is a persistent, low luminosity
pulsar in a binary orbit with the Be star X~Persei (X~Per). Its
\wsim837\,s pulsation period was discovered with the \uhuru\
satellite \citep{whi76,whi77}, and is still one of the longest
periods of any known accreting pulsar \citep[][and references
therein]{bil97}. The distance to the system, based on optical
observations of the companion X~Per, is 0.95\wpm0.20\,kpc
\citep{tel98}. Recently, \citet{del00} have determined a complete
orbital ephemeris of the system using data from the \RXTE\
(\rxte).

Like many accreting pulsars, the history of the pulse period of
4U~0352+309 shows both episodes of spin-up and spin-down
\citep[][and references therein]{whi76,rob96,sal98}. The neutron
star experienced a large apparent torque reversal around 1978.
4U~1626$-$67 \citep{cha97} and GX~1+4 \citep{mak88} have
exhibited similar torque reversals, and in both sources the mean
spin-up and spin-down rates were similar in absolute magnitude.
Previous to 1978 the spin history of 4U~0352+309 was erratic;
however, there was a general trend toward spin-up at a rate of
$\dot{P}_{\rm spin}/P_{\rm spin} = -1.5\dx{-4}$\,yr$^{-1}$
\citep{del00}. Since 1978, although the spin period measurements
have been infrequent the star has been spinning down at an average
rate of $\dot{P}_{\rm spin}/P_{\rm spin} = 1.3\dx{-4}$\,yr$^{-1}$
\citep{del00}. Unfortunately there are not enough spin period
measurements to distinguish between steady spin down, as might be
due to disk accretion, or the erratic torque fluctuations that
would be expected from wind accretion.

An unusual aspect of 4U~0352+309 is the source's \xray\ spectrum.
Most accreting \xray\ pulsar spectra are well described by a
standard model: a simple power-law with a photon index of \wsim1
that is exponentially cutoff above \wsim20\,keV \citep{whi83}.
4U~0352+309, on the other hand, does not follow this standard
spectral shape and as a result has been fit with a variety of
different models. Observations lacking response below \wsim2\,keV
have typically been fit with thin thermal bremsstrahlung emission
in the temperature range 7--12\,keV \citep{bec79,whi76,whi82} or
thermal bremsstrahlung with a high energy tail
\citep{mus77,fro79,wor81,whi82,fro85}. Observations obtained with
the \tenma\ \citep{mur87}, \exosat\ \citep{rob89}, and \ginga\
\citep{rob96} satellites were fit by the standard model for binary
\xray\ pulsars, but required very low cutoff energies (in the
range of 0.6--1.7\,keV) that are atypical of most accreting
pulsars \citep{whi83}. Furthermore the \ginga\ result, where the
data extended to higher energies, showed evidence for a hard tail
above \wsim15\,keV. Recently \citet{sal98} fit the broad band
(0.1--200\,keV) \sax\ \xray\ spectrum with a combination of two
power-laws. The first, which was important at lower energies, had
a simple high-energy cutoff and was identical in form to the
standard spectral model for accreting pulsars. The second
power-law, which dominated above \wsim15\,keV, had both high and
low-energy cutoffs.

With the broad energy ranges and good spectral resolution of
satellites such as the \rxte\ and \sax, several new \CRSFs\
(\crsfs) have been discovered in well known accreting pulsars.
Two of the most notable recent discoveries are Cen~X$-$3
\citep{san98,biff99} and 4U~1626$-$67 \citep{orl98,biff99}. These
line-like spectral features are thought to be due to resonant
scattering by electrons in Landau orbits in the \wsim\ex{12}\,G
magnetic field of the neutron star. The \crsf\ centroid energy is
given by \crsfeq{}\,keV, where $z$ is the gravitational redshift
in the scattering region and $B_{12}$ is the magnetic field in
units of \ex{12} Gauss. Thus \crsfs\ give a direct measurement of
the average magnetic field in the scattering region near the
neutron star surface. Without a \crsf\ the magnetic dipole moment
of the star can only be inferred from measurements of the
luminosity (a function of the often uncertain distance) and spin
period using accretion torque theory \citep{gho79}. Even the
radio pulsar magnetic fields are estimated, not directly
measured, assuming dipole radiation and measuring the period and
period derivative \citep{manchester77}.


\section{Observations and Analysis}\label{sec:obs}

The observations of 4U~0352+309 studied here were made using the
two pointed instruments on the \rxte. The \PCA\
\citep[\pca,][]{jah96} is a set of five identical Xenon
proportional counters (\PCUS, or \pcus) sensitive in the energy
range 2--60\,keV. The \pca\ can, and sometimes does, operate with
one or more of the individual \pcus\ turned off. This is done to
extend the lifetime of the instrument, but reduces the effective
area during those observations. The \HEXTE\
\citep[\hexte,][]{rot98} consists of two clusters of 4
NaI(Tl)/CsI(Na) phoswich scintillation detectors (15--250\,keV)
that rock on and off source, simultaneously measuring both source
and background counts. One of the detectors in Cluster B suffered
a failed pulse height analyzer early in the mission and no longer
provides any spectral information, effectively reducing the area
of that cluster by 25\%. The \pca\ and \hexte\ fields of view are
co-aligned on-source and are collimated to the same 1\degree\
full width half maximum (FWHM) region.

\begin{figure}[ht]
\centerline{\includegraphics[]{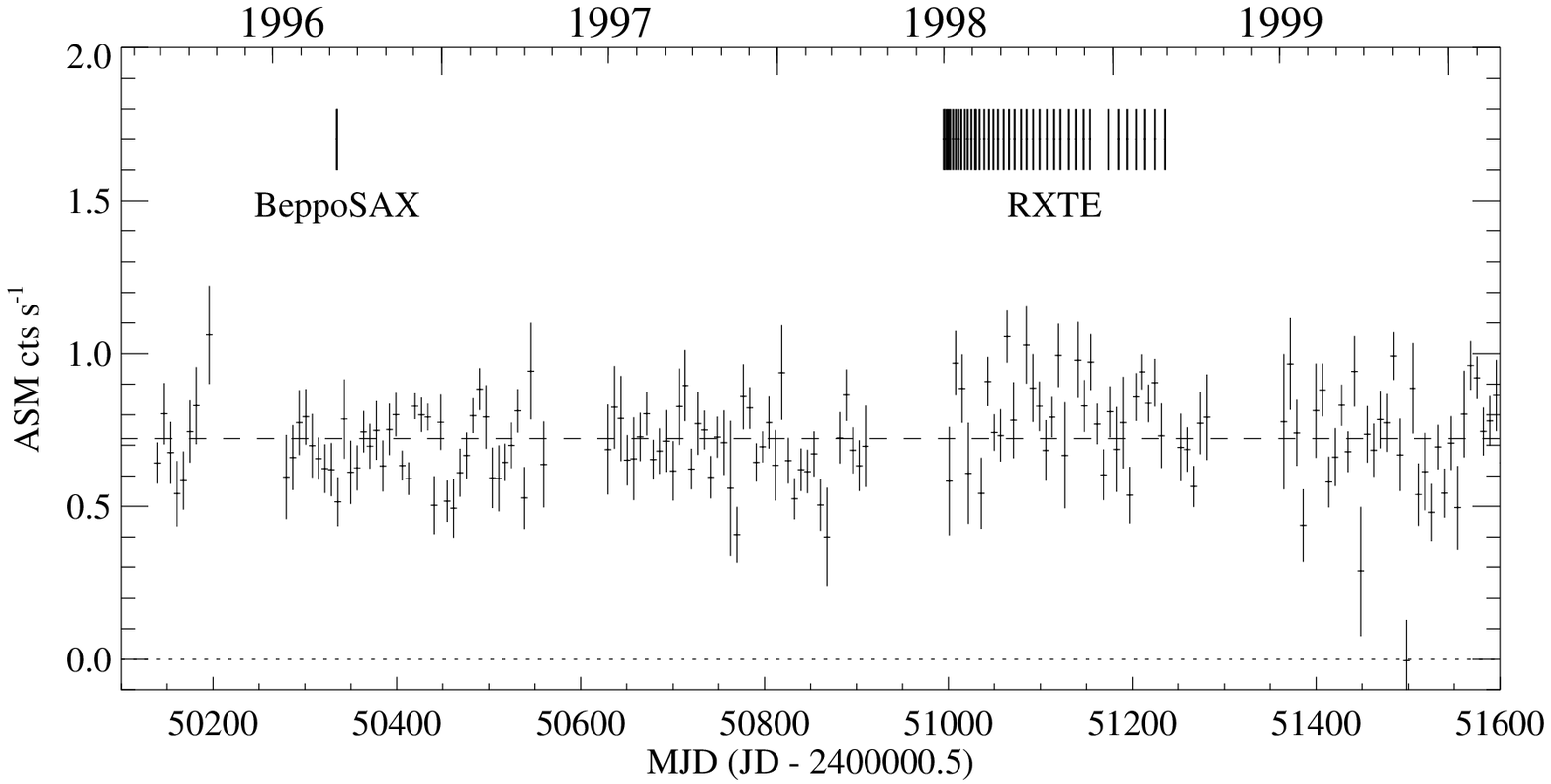}}
\caption{\label{fig:asm} Counting rate of 4U~0352+309 in the \asm,
binned as weekly averages. The thick bars near the top of the plot
indicate where observations were made with the \rxte\ pointed
instruments, as well as the \sax\ observation \citep{sal98} for
comparison. The average \asm\ counting rate (0.72\wpm0.1\,cps) is
shown as a dashed line. The average rate during the period of
\rxte\ pointed observations is 0.82\wpm0.02\,\cps, above the
overall source average. For comparison, the Crab Pulsar is
\wsim75 cts/sec in the \asm.}
\end{figure}

We have used data from 40 pointings from the \rxte\ public archive
(see Table~\ref{table:obs}) spanning from 1998 July 1 through 1999
February 27, roughly one binary orbit of the system. This is a
subset of the observations used by \citet{del00}, allowing us to
use their orbital ephemeris in our timing analysis. These
observations were chosen for their relatively short intervals
between pointings (on average 6 days), which minimizes the
possibility of long-term spectral variability and yet provides
enough data for a suitable detection of the source. See
Fig.~\ref{fig:asm} for the \rxte/\asm\ history of the source,
along with the times of the \rxte\ pointings.

For the \pca\ we used GoodXenon mode data with \pcus\ 0, 1, and 2
only, as these were the only \pca\ detectors that were on during
all of the observations. The total \pca\ on-source exposure was
161\,ks with 3 \pcus, and the on-source livetime in \hexte\ was
54.6\,ks per cluster. The average \pca\ 3--25\,keV counting rate
was 20.6\wpm0.1\,\cpspcu. In \hexte\ the counting rate was
1.78\wpm0.06\,\cps\ in Cluster A and 1.30\wpm0.05\,\cps\ in
Cluster B. The average 3--100\,keV \xray\ flux from the source
over the course of the observations was 3.9\dx{-10}\,\flux. At a
distance of 0.95\,kpc \citep{tel98} the source luminosity is
4.2\dx{34}\,\lum. This is at least two orders of magnitude less
than the typical \ex{36}-\ex{38}\,\lum\ luminosities seen in
other persistent accreting pulsars or transient pulsars while in
outburst.

The phase averaged \pca\ and \hexte\ spectra were accumulated
using version 5.0.1 of the HEASoft tools. Even though the average
\pca\ counting rate was below the 40\,\cpspcu\ threshold for
using the Faint source background model, the large pulse
modulation often brought the instantaneous counting well above,
at times over 100\,\cpspcu. Since we cannot, at this time, phase
resolve the background spectra of the \pca, we were forced to use
the bright-source SkyVLE model. The Faint source model, which
relies partly on average counting rates, significantly
over-predicted the background level. In particular, the modeled
Faint source background was larger than the measured source plus
background for the phase minimum at all energies. The SkyVLE
model, on the other hand, predicted a background rate that was
less than source plus background for all four phases. These
models also gave background subtractions that were consistent
with zero at high energies, where 4U~0352+309 was too weak to
detect. Therefore we concluded that it was appropriate to use the
the SkyVLE models for our \pca\ background subtraction. Since the
\hexte\ uses a measured background accumulated in the same way as
the source pointings, it did not suffer these problems. The
estimated error of the \hexte\ background subtraction is a few
tenths of a percent \citep{rot98}, while the source flux of
4U~0352+309 is 1.7\% of background.

In order to study the evolution of the spectrum through the
pulse, and to aid in the search for \crsfs, we performed pulse
phase resolved spectral analysis as well. The spectra of
accreting \xray\ pulsars can vary considerably with pulse phase
\citep{whi83}. This is easily seen in the Folded Light Curves
(\flc s) of these sources, where the general trend is toward more
complicated pulse shapes at lower energies, and larger pulsed
fractions at higher energies. This two dimensional nature can
make interpreting spectra averaged over an entire pulse period
problematic.

\begin{figure}[ht]
\centerline{\includegraphics[]{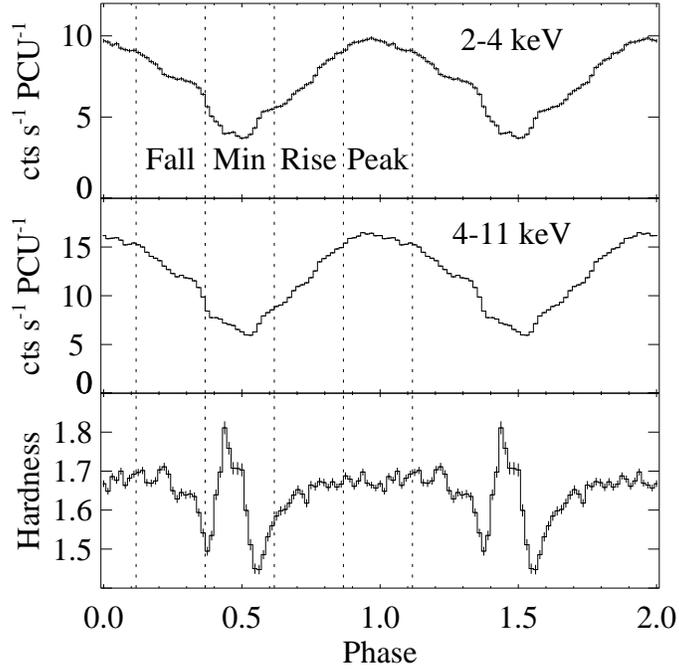}}
\caption{\label{fig:pca_hardness} TOP: \pca\ folded light curve in
the 2-4\,keV range. The phase intervals used in the spectral
analysis (Peak, Fall, Min, and Rise) are indicated by the dotted
lines. MIDDLE: Folded light curve in the 4-11\,keV range. BOTTOM:
Hardness ratio, defined as ratio of counts in the 4-11\,keV band
to those in 2-4\,keV. As in previous observations
\cite{rob89,rob96} there is a prominent peak in the hardness
during the pulse minimum. This indicates that the average \xray\
spectrum of 4U~0352+309 is relatively constant. We do note,
however, that when the \sax\ satellite observed the source
\citep{sal98} the hardness spike was much less prominent.}
\end{figure}

We corrected photon arrival times to both the solar system and the
binary system barycenters using the ephemeris of \citet{del00}
and accumulated spectra using a locally modified version of the
\fasebin\ phase resolved accumulation FTOOL. The \fasebin\ FTOOL
produces a two dimensional counts histogram versus both phase and
energy, allowing for an analysis of spectra as a function of
phase, or folded light curves as a function of energy.
Modifications to the \fasebin\ FTOOL were made to take the
\hexte\ deadtime corrections into account properly and to allow
for phase resolving the \hexte\ background files. Phase resolving
the background is important for long period pulsars, where the
background level can vary over the course of a single pulse.
Given the relatively simple, saw-tooth like pulse shape of
4U~0352+309 and the faintness of the source, we divided the pulse
into four phase bins; Peak, Fall, Min, and Rise (See
Fig.~\ref{fig:pca_hardness}).

The uncertainties in the \pca\ spectra, with its large collecting
area and thus excellent counting statistics, are dominated by the
systematic errors in the response matrix and background modeling
\citep{wil99}. To estimate the size of the systematic errors in
the \pca\ version 2.43 response matrix, we analyzed observations
of the Crab nebula and pulsar as recommended by the \pca\ team
\citep{jah00priv}. To be able to compare directly to our
4U~0352+309 composite spectrum, we summed 10 Crab observations
taken between 1998 July 27 and 1999 February 24, for a total \pca\
exposure of 8\,ks. To fit these data, we used a combination of two
power-laws. The first power-law, with a best fit photon index of
$2.21_{-0.02}^{+0.03}$, was to account for the nebular flux. The
second, which modeled emission from the pulsar, had a photon
index of $1.9_{-0.1}^{+0.7}$ and a 2--10\,keV normalization fixed
to be 10\% of the first \citep{jah00,kni82}. We then examined the
best fit and increased the size of the systematic errors as a
function of energy until we achieved a reduced \wchi\ of unity.
The second power-law is important in this type of analysis, since
ignoring the contribution of the pulsar would have led us to
estimate larger systematic errors.

The systematic errors we used were 0.4\% in the range
2--12.5\,keV, 0.3\% in 12.5--20.5\,keV, and 2.4\% above 20.5\,keV.
These are slightly smaller than the errors used by other authors
\citep[eg.][]{wil99,bar00}, due in part to the dual power-law
model used and improvements in the current response matrices
(PCARSP V2.43). The effects of increasing the systematic errors
reduces the formal significance of the \crsf\ by improving \wchi,
as well as increasing the uncertainties of the fit parameters.
However, such an increase does not affect the overall shape of
the residual errors, especially the line-like shape in the \hexte\
data around 30\,keV. The errors of the \hexte\ data are dominated
by the counting statistics \citep{rot98}, so systematic errors
were not applied.

In our analysis of 4U~0352+309 we truncated the \pca\ energy range
at 25\,keV. This was done to minimize the effects of the large
systematic errors above 20\,keV, while still allowing a
significant overlap between the \pca\ and \hexte\ energy ranges.
We fit the \pca\ simultaneously with data from both \hexte\
clusters in the range 16--100\,keV, with the overall
normalization of each \hexte\ cluster allowed to vary with
respect to the \pca\ in the fits.

\subsection{Spectral Variability}\label{sec:specvar}

One of our major concerns was the validity of summing a large
number of short pointings, taken over the course of 8 months, into
a single set of spectra. This is especially dangerous when
searching for \crsfs, as averaging different power-law slopes
could artificially create a dip-like or inflection feature in the
spectrum. The pulse profile of 4U~0352+309 is also known to vary
from pulse to pulse \citep{del00}, and due to the long pulse
period there are fewer than 6 pulses in any given 5\,ks
observation. Furthermore, the \xray\ spectra of accretion powered
pulsars, especially wind fed systems, can vary dramatically with
both time and luminosity \citep{kre99}. We took great care to
ensure that the spectra we analyzed were free from these effects.

Since the separation of 4U 0352+309 and X~Per is quite large
\citep[2.2\,AU,][]{del00}, the system is almost certainly wind
fed. But because of the modest eccentricity
\citep[e=0.11,][]{del00}, we did not expect the effects of the
stellar wind to vary dramatically with the orbit. Furthermore, the
small inferred inclination angle
\citep[$i\sim23-30$\degree,][]{del00} implies that the system is
not being viewed through a (possibly variable) stellar or
accretion disk. Because of this we did not expect significant
variations in the absorption column, and thus in low energy
spectral shape, over time. Additionally, although no single
observation was long enough to make a significant detection of a
\crsf, the longest single pointing suggested that there might
indeed be one. Therefore we were motivated to investigate whether
we could discuss the average spectrum of 4U~0352+309.

First we tested whether or not the pulse shape converged to an
average, repeatable shape. We divided the 40 pointings into 5
contiguous sets of observations, with each segment containing
approximately 38 pulses. We then did a two component fit of each
of the five 3--25\,keV \flc s to the average \flc, allowing for
both an additive offset in rate and an overall multiplicative
factor. This allowed for possible errors in the background
subtraction and tested the repeatability of the overall shape,
but was insensitive to changes in pulsed fraction. The fit
offsets were small, only 3--4\% of the rate in each phase bin.
The scatter around the mean \flc\ gave reduced \wchi s of 1.05,
0.66, 0.59, 0.67, and 0.91 in the 5 data segments. So, as with
other accreting pulsars, the pulse shape variations of 4U~0352+309
converge to an average pulse shape that is stable at least over
several months.

Next we sought to verify whether the average pulse was similar to
previous observations. In particular 4U~0352+309 has shown a peak
in hardness at pulse minimum \citep{rob89,rob96,sal98}. Using the
spectra from our phase resolved analysis we generated folded
light curves in the 2--4 and 4--11\,keV bands and created a
hardness ratio through the pulse (Fig.~\ref{fig:pca_hardness}).
As in previous observations, we see a nearly constant hardness
ratio though the peak of the pulse as well as a hard spike at
pulse minimum. This consistent behavior suggests that summing a
series of short observations would be nearly equivalent to a
single long pointing. We also note that the hard spike is very
prominent in our data, similar to what was observed with \exosat\
\citep{rob89} and \ginga\ \citep{rob96}, while the spike was much
less pronounced during the \sax\ observation \citep{sal98}.

Lastly, we investigated whether or not the spectral shapes in the
five data segments were consistent with the average spectral
shape. To account for the fact that the 5 intervals had different
amounts of time in different parts of the pulse, we used
background subtracted spectra that were also resolved into our 4
phase bins (see above), giving a total of 20 spectra to be
tested. Then, for the 5 segments in each phase bin, we took the
ratio of the 3--25\,keV \pca\ counts to that of the total counts
spectrum (also background subtracted) for that phase bin, and
calculated the reduced \wchi\ assuming a constant ratio. This was
done for all 4 phases. See Table~\ref{table:5segfits} for the
results of this analysis. We found that the spectral shape in the
5 data segments, at least in the 3--25\,keV range, is consistent
with being constant. Therefore, due to the stability of the pulse
and spectral shapes over the course of our observations, we
concluded that summing data from all of the observations would
not introduce a significant artifact in a given phase bin or for
the phase averaged spectrum.

\subsection{Spectral Fitting}\label{sec:specfit}

\begin{figure}[ht]
\centerline{\includegraphics[]{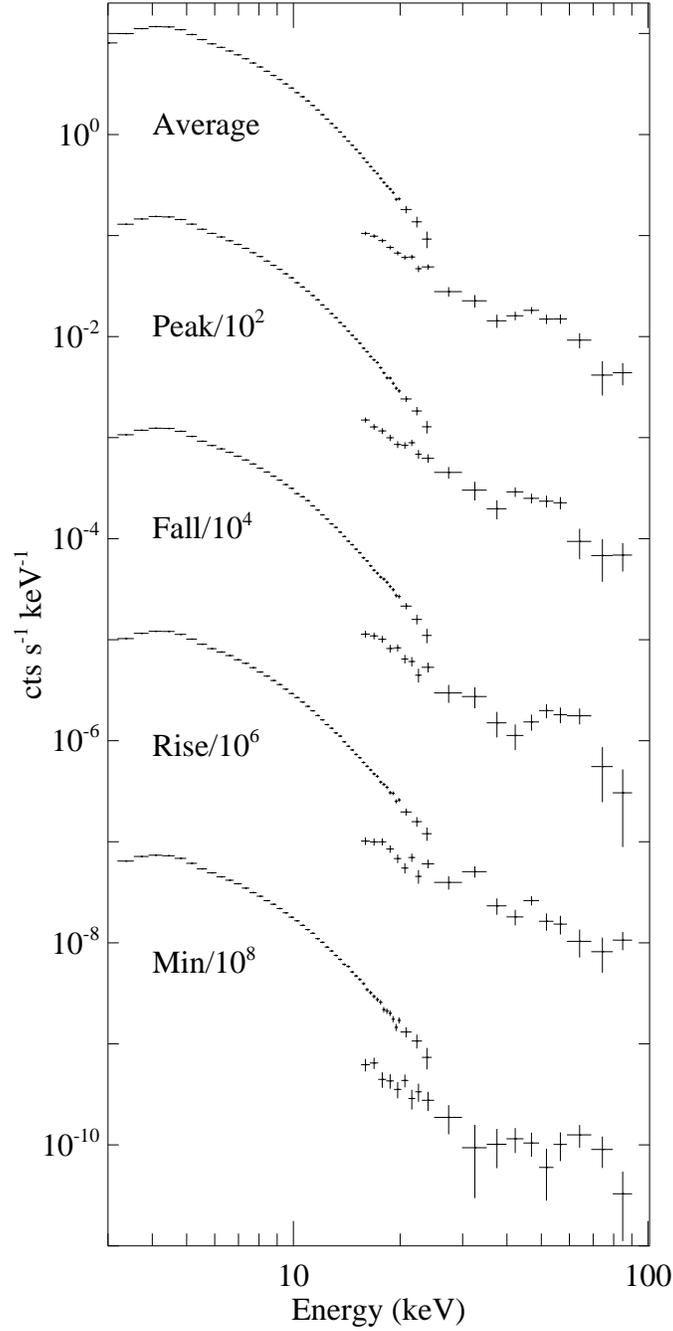}}
\caption{\label{fig:counts} Plot of the raw \pca\ and \hexte\
counts for the phase average spectrum and four phase bins defined
in Fig.~\ref{fig:pca_hardness}. The \pca\ is shown from 3-25\,keV
while the \hexte\ is 16-100\,keV. The \crsf\ is visible as in
inflection in the phase average counts, and especially so in the
falling edge. }
\end{figure}

Identifying \crsfs\ in the spectrum of accreting pulsars can be
problematic, especially when the underlying continuum is unknown.
However, in the case of 4U~0352+309 a large feature is clearly
visible at \wsim50\,keV in the counts spectra of the source
(Fig.~\ref{fig:counts}), and is especially evident in the falling
edge of the pulse peak. This same feature is also visible in a
shorter observation (\wsim10\,ks) made early on in the \rxte\
mission. This feature is not present in counts spectra
successfully fit with the standard model of accreting pulsars,
but would be expected if there were also resonant cyclotron
scattering present in the source. We were therefore motivated to
find a simple and physically reasonable continuum spectrum that
would describe the data.

In our fits we fixed the neutral hydrogen column to be $1.5
\dx{21}$\pcmsq, based on measurements from satellites with
significant response below 1\,keV (for example \rosat,
\citealt{hab94,mav93}, \bbxrt, \citealt{sch93}, \copernicus,
\citealt{mas76}, and \sax, \citealt{sal98}). This is an order of
magnitude smaller than measurements made by instruments lacking
response below \wsim2\,keV \citep[e.g.][]{whi82,mur87,rob96}. It
is difficult for these instruments (e.g. \ginga\ and \rxte) to
measure such a low column. The fitting is also complicated by the
fact that there is a change or break in spectrum at \wsim2\,keV,
as revealed by instruments such as \exosat\ \citep{rob89} and
\bbxrt\ \citep{sch93}. This change can explain why instruments
without low energy response systematically measure a column which
is larger by an order of magnitude.

Starting with models used historically for 4U~0352+309 (see
\S~\ref{sec:intro}), we first attempted to fit the \rxte\ data
with a thermal bremsstrahlung model both with and without a
power-law tail. We found that these models gave poor fits to the
data when the absorption column was fixed at 1.5\dx{21}\,\pcmsq.
When the amount of absorption was allowed to vary we achieved
acceptable fits, but these required an unreasonably large column
($1-3\dx{22}$\,\pcmsq) and were consistent with previous
observations where a thin thermal plasma model has been applied to
$>2$\,keV data. The thermal model also suffers from no iron line
ever being detected in this source (with an upper limit on the
equivalent width of \wsim6\,eV, \citealt{sal98}). Because of this
and the large inferred absorption column we have rejected the
bremsstrahlung and bremsstrahlung plus power law tail continuum
models.

In the \rxte\ band, the standard model used to describe accreting
\xray\ pulsars is a power-law that breaks into a power-law times
an exponential at a characteristic cutoff energy. Historically
this has been realized in the \plcut\ form \[\plcuteq\] where
\wgamma\ is the photon power-law index, and \ecut\ and \efold\
are the cutoff and folding energies respectively. Other
analytical forms have been used as well, such as the Fermi-Dirac
form of the cutoff \citep[\fdco,][]{tan86} and the \NPEX\ model
\citep[\npex,][]{mih95}. For a full review of the various
phenomenological models see \citet{kre99}. This generic spectral
shape can be interpreted as due to Comptonization of soft photons
by the hot accreting electrons, with the cutoff and folding
energies determined by the temperature of the infalling electrons.

When we applied these models (\plcut, \fdco, and \npex) we found
that they were inadequate to describe the shape of the spectra.
As with \ginga\ \citep{rob96}, there appeared to be a power-law
tail extending beyond \wsim15\,keV. Our inferred values of cutoff
energy were below 3\,keV (and outside the \pca\ energy range),
also consistent with previous results. Moreover, the ``tail'' at
higher energies was still present when we fit the pulse peak
minus pulse minimum, confirming that it is pulsed emission from
4U~0352+309 and not a contaminating background source such as an
AGN. However, the simple addition of an extra power-law to the
model, or even a second \plcut, still gave unacceptable fits.
Specifically these were unable to account for the dip seen at
\wsim35\,keV. This motivated us to try other two component models.

Di Salvo et al. (1998) fit the \sax\ spectrum of 4U~0352+309 with
a dual \plcut\ model. The first \plcut\ had a cutoff energy of
\wsim2\,keV. The second, which was only important above
\wsim\,15keV where there was ``extra'' emission, had a cutoff
energy of 66\,keV and an exponentially shaped low-energy
absorption term with a 44\,keV folding energy. The cross over of
the two components was at \wsim15\,keV, and the second component
fits the excess ``tail'' not described by the lower energy
component. We applied this model to our observations as well and
achieved a reduced \wchi\ of 1.25 for 79 degrees of freedom for
the phase averaged spectrum. The shape of the spectrum, however,
was noticeably different from that obtained by \citet{sal98}. In
particular, the high energy photon index we obtained was smaller
(although within errors, 1.5\wpm0.6 vs. 2.6\wpm1.1) and the low
energy exponential absorption folding energy an order of
magnitude less ($2.7_{-2.2}^{+3.3}$ vs. 44.3\wpm4.5\,keV).
Furthermore, the \rxte\ fits using this model were improved by
the addition of a \crsf\ at \wsim30\,keV, giving a reduced \wchi
of 0.753 for 76 degrees of freedom. This difference between the
spectra obtained with the \sax\ and the \rxte\ satellites could
be real, possibly connected with variations of the source
luminosity. The 1--10\,keV flux during the \sax\ observation was
1.7\dx{-10}\,\flux\ \citep{sal98}, while the 1.7--10\,keV flux in
our observation was 2.1\dx{-10}\,\flux, or 23\% higher.
Furthermore, the \rxte/\asm\ source flux in the week spanning the
\sax\ observation was 0.52\wpm0.08\,\cps, which is below the
source average (Fig.~\ref{fig:asm}).

\citet{nel95} suggested that the spectrum of accreting
low-luminosity neutron stars ($L_{\rm X} \lesssim \dx{34}$\,\flux)
could exhibit a cyclotron \emph{emission} line. This feature is
expected to be broad ($E/\Delta E \sim 2-4$), peak at energies
below the fundamental cyclotron energy of the magnetic field, and
be sharply cutoff above the cyclotron energy. Di Salvo et al.
(1998) interpreted the second \plcut\ in their spectrum as
evidence for this feature. Since the second component in our fits
was different from that obtained with the \sax, and quite unlike
an emission line, we attempted to test the \citet{nel95}
prediction with some slightly different models. We fit the phase
averaged spectrum with the standard \plcut\ and \fdco\ forms,
plus an additional Gaussian shaped emission line, both with and
without a high-energy cutoff. The Gaussian was meant to fit the
peak near 50\,keV, while the cutoff was used to truncate the
high-energy wing of the Gaussian to simulate the predicted steep
fall at the cyclotron energy. The folding energy of the cutoff
associated with the Gaussian component was allowed to vary, but
constrained to be within 0.1--10\,keV. We found that the presence
of a cutoff didn't change the fits significantly, and we were
unable to achieve any reasonable fits using an emission line
model.

\begin{figure}[ht]
\centerline{\includegraphics[]{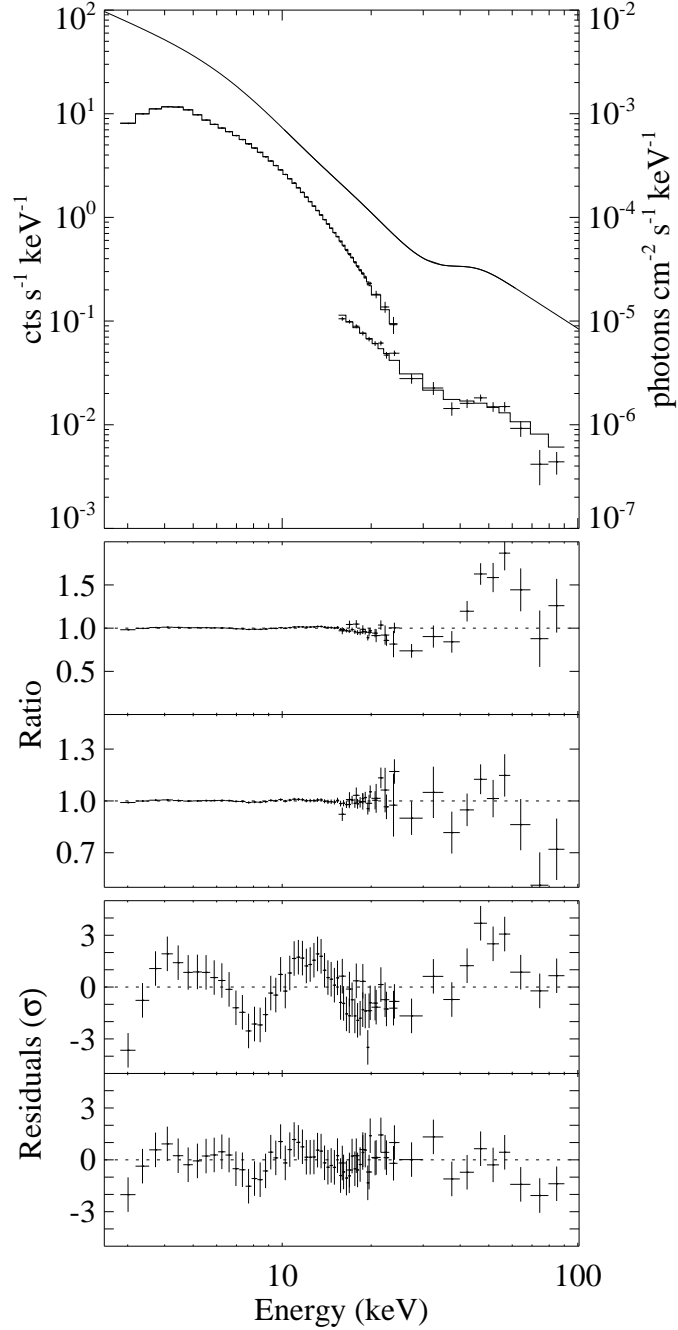}}
\caption{\label{fig:phavg} Top: Plot of the phase averaged counts
spectra (crosses) fit with a combination of a black body and power
law (histograms). The input model is also shown (smooth line).
Middle: Ratio of data to model without (top) and with (bottom) a
\crsf\ at \wsim29\,keV. Bottom: The residuals to the fit, in
units of sigma, both without (top) and with (bottom) a
\wsim29\,keV \crsf. The \pca\ points contain the systematic
errors discussed in the text (see \S~\ref{sec:obs}) }
\end{figure}

We have found a two component model that both best described the
data and had a simple physical interpretation. The first
component was a \wsim1.4\,keV black-body, the second a power-law
with a photon index \wgamma\wsim1.8 and a broad absorption
feature centered at 29\,keV. See Table~\ref{table:bbodyfits} for
the best fit parameters and Fig.~\ref{fig:phavg} for the phase
averaged spectrum.

\begin{figure}[ht]
\centerline{\includegraphics[]{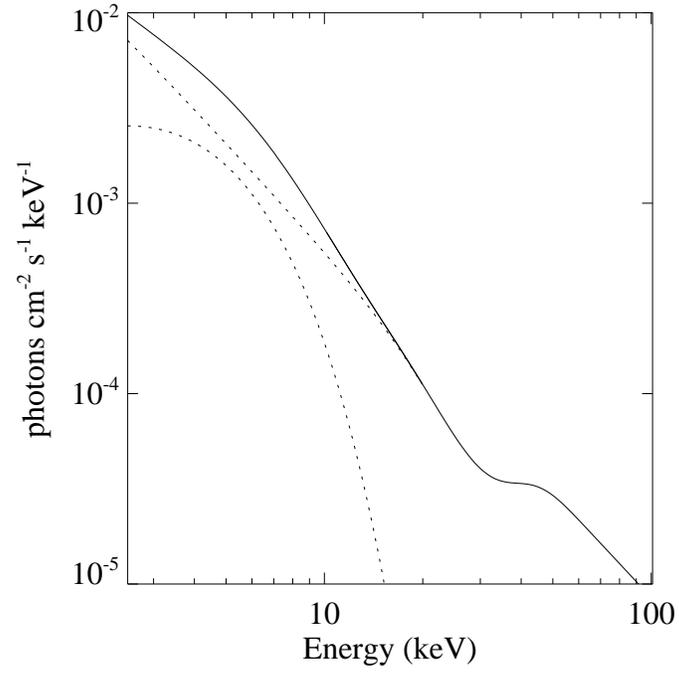}}
\caption{\label{fig:model} The inferred model from the phase
averaged spectral fits, along with the black body and power-law
components show individually as dotted lines. The \crsf\ is
visible as a notch in the power-law. }
\end{figure}

As in previous observations of the source we found no evidence of
an Fe-K line in the spectrum, with a 90\% confidence upper limit
on the equivalent width of 13\,eV for a Gaussian line centered at
6.4\,keV and with a 0.5\,keV sigma. This is consistent with the
\wsim6\,eV upper limit of \citet{sal98}. In the phase average
fits, the inferred area of the black body was $5.4\dx{8}$\,\cmsq,
consistent with standard values for the magnetic polar cap of a
neutron star. The black-body temperature remained constant, to
within errors, with pulse phase. We note, however, that the
formal errors on the black-body component are small (only 20\,eV
for the temperature). Factors such as the fixed value for the
absorption column and the spectrum below 1\,keV, which is not
measured with the \rxte\ and was fit with a slightly different
shape in the \sax\ data\citep{sal98}, affect these results.
Therefore the actual errors for both the temperature and area are
almost certainly larger. See Fig.~\ref{fig:model} for the two
model components in the inferred phase average photon spectrum.

\begin{figure}[ht]
\centerline{\includegraphics[height=5in]{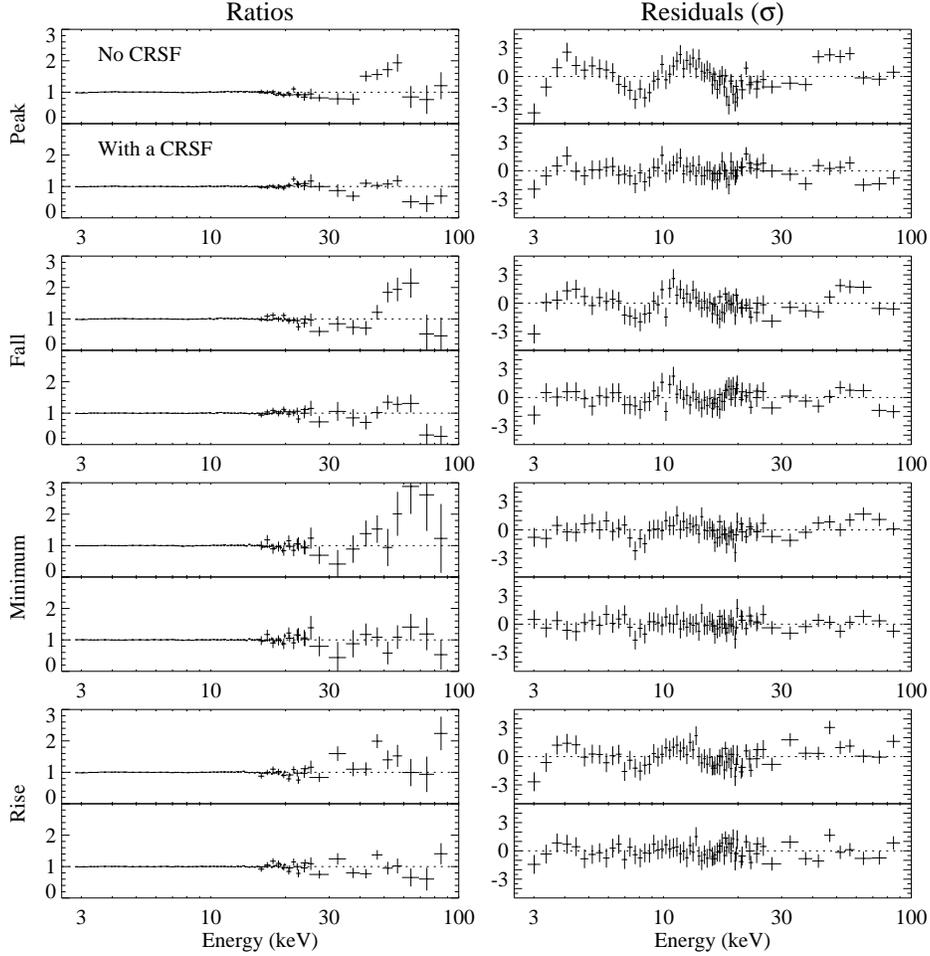}}
\caption{\label{fig:ratios} The ratios (left column) and residuals
(right column) of the data to the best fit model before (top
panels) and after (bottom panels) the addition of a \crsf\ to the
model, in the four phase bins. In addition to errors due to the
counting statistics, the \pca\ points contain the systematic
errors discussed in the text (see \S~\ref{sec:obs}). Due to the
falling nature of the continuum, there are more counts at
energies below the line centroid, and therefore these data have a
larger effect on the fitting procedure. When fit without an
absorption line this leads to a continuum that hugs the lower
edge of the line and an under prediction at energies above the
line. In a ratio plot this manifests itself as a dip followed by
an excess, and is a classic signature of a \crsf. This signature
is especially evident in the pulse fall and minimum. Although we
do not find strong evidence for a spectral cutoff (see
\S~\ref{sec:discuss}), the highest few \hexte\ energy bins in
each of the panels are an indication that there might be a
high-energy cutoff.}
\end{figure}

The absorption line model used in the fits was a multiplicative
exponential of the form $\rm{e}^{-\tau(E)}$, where $\tau$ is the
optical depth of the absorption as a function of energy. The
functional form of $\tau$ is given by the Gaussian \gabseq{} where
\ecy{} is the centroid energy (also called line energy), \wcy{}
is the width, and \dcy{}\ is the optical depth at the \crsf\
energy. We interpret this feature as a 29\,keV \crsf. It was
present in all four phases; however, it was only weakly detected
during the rise of the pulse. In Fig.~\ref{fig:ratios} we plot
the ratio of the data to the best fit model, both before and
after the addition of a \crsf\ to the model. Because there are
simply more counts at energies below the centroid energy, the
errors are smaller and have a greater effect on the fitting
process. When fitting without a \crsf, these smaller errors drive
the continuum to fit the low side of the line and under predict
the continuum at higher energies. This gives rise to a dip
followed by a rise in a ratio plot -- a classic signature of a
\crsf. These dips are especially evident in the pulse fall and
minimum in Fig.~\ref{fig:ratios}.

The \crsf\ does, however, appear near the 33.17\,keV K-edge of
Iodine, where there is a dramatic change in the response
properties in the \hexte\ NaI scintillators. In a ratio plot
(Fig.~\ref{fig:phavg}), the trough of the feature is below
30\,keV, and shows a deviation from the model of approximately
20\%. There is a similar structure in the \hexte\ ratios of
power-law fits to the Crab nebula/pulsar using the released
matrices. This structure, however, is significantly smaller (less
than 1\%) and is centered above 30\,keV. Additionally, when we
take the ratio of the 4U~0352+309 counts spectrum to that of the
Crab, we still see a line-like structure. Since the spectrum of
4U~0352+309 is a power-law (with the exception of the \crsf) in
the \hexte\ band, this ratio is, to first order, free of
instrumental effects. Lastly, we fit the data using a set of
intermediate response matrices provided by the \hexte\ instrument
team. The released version of the \hexte\ response matrices are
derived from these, but have had the diagonal elements adjusted
slightly to give a smooth fit to the Crab and account for
calibration errors near the K-edge. The intermediate matrices,
which are based solely upon modeling of the detectors and
calibration data, have residual errors that are \wpm2\% in the
region of the K-edge. When we fit our 4U~0352+309 spectrum with
these unreleased matrices \wchi\ becomes worse; however, the line
width and energy remain unchanged. Due of this analysis we
conclude that the feature is intrinsic to the source, and not due
to any instrumental problems or artifacts. We further conclude
that the \crsf\ fit parameters are not significantly affected by
the instrumental calibration.

From the pulse peak through minimum the \crsf\ energy is constant
to within errors and is quite broad. The fit parameters are also
consistent with the phase averaged spectrum. At pulse minimum a
feature can be clearly seen in the pattern of ratios, although
the statistics in this phase bin are quite low. The addition of a
\crsf\ is not formally required by the fits, since a reduced
\wchi\ of 0.96 can be achieved without it. However, the addition
of a \crsf\ at 30\wpm6\,keV greatly improves the fits, with an
F-test chance probability of $3.5\dx{-5}$.

The fit \crsf\ during the pulse rise is quite different than the
other phases. Specifically, it appears at a lower energy than in
the other three phases (21 instead of \wsim29\,keV) and with a
shallower depth. The \crsf\ also falls in the region of overlap
between the \pca\ and \hexte\ instruments, and due to its width
is not completely resolved by either instrument. Because of this
the fitted \crsf\ might be an artifact. We do note, however, that
the width of the \crsf\ is, as a percentage of the centroid value,
similar to the other three phases, and the temperature of the soft
component and power-law index are consistent with the other pulse
phases.

We also conducted a search for a \crsf\ second harmonic. Multiple
harmonics have been detected in sources such as 4U~0115+63
\citep{hei99,san99} and 4U~1907+09 \citep{cus98}, and so it is not
unreasonable to search for one in this source as well. There
appears, however, to be no evidence for a higher harmonic at
\wsim60\,keV. Our statistics in that range are poor and resolving
a line would be difficult. Using the phase average spectra we
added a \crsf\ fixed at twice the energy and width of the
fundamental. We then increased the depth parameter until the
\wchi\ changed by 2.7 (for one interesting parameter), and were
able to obtain a 90\% confidence upper limit of 0.25 on the depth
of a second harmonic.


\section{Discussion}\label{sec:discuss}

Since the discovery of the first cyclotron resonance scattering
feature in the spectrum of Her~X-1 \citep{tru78}, there have been
only about a dozen firm detections of \crsfs\ out of the \wsim50
known accreting \xray\ pulsars \citep{mak99}. With this analysis
we have added another, namely 4U~0352+309. The magnetic field at
the polar cap implied by a 29\,keV \crsf\ is $2.5(1+z)\dx{12}$\,G
(where $z$ is the gravitational redshift in the scattering
region). Assuming the radius and mass of the neutron star to be
10\,km and 1.4\,\msol\ respectively, and that the cyclotron
scattering is occurring at the magnetic polar cap, then the
magnetic field is $3.3\dx{12}$\,G. This field strength is well
within the range of other accreting pulsars as measured by \crsfs,
and also near the $2\dx{12}$\,G peak in the distribution of radio
pulsar fields \citep{tay93}. This implies that there is nothing
unusual about the magnetic field strength of 4U~0352+309.

It is often useful to discuss the magnetic dipole moment $\mu$ of
the neutron star since the magnetic field strength is a strong
function of position. For a purely dipolar field, the dipole
moment is given by $\mu = 0.5\,B\,R^{3}$, where $B$ is the
magnetic field strength at the polar cap and $R$ the radius of
the neutron star \citep{shapiro83,jackson75}. So, for a polar cap
field of 3.3\dx{12}\,G, the magnetic dipole moment of 4U~0352+309
is 1.7\dx{30}\,\mom.

In our two component continuum, we identify the power-law as the
normal accreting pulsar continuum shape in the low accretion rate
limit. While the spectrum of most accreting pulsars are
exponentially cutoff above a characteristic energy \ecut\
\citep{whi83}, our fits of 4U~0352+309 did not strongly require
such a cutoff. When we substituted a \plcut\ for the simple
power-law in our model (see \S~\ref{sec:specfit}), the cutoff and
folding energies were $\ecut = 57_{-17}^{+12}$ and $\efold =
50_{-30}^{+107}$ respectively, with an F-test chance probability
of 6\dx{-3}. This is consistent with the results of \citet{sal98}
when fitting \sax\ data: $\ecut = 66_{-14}^{+28}$ and $\efold =
44_{-22}^{+54}$. However, when we used the slightly different
\fdco, the cutoff and folding energies were completely
unconstrained. Because of the large errors and model dependence,
we find there is at most weak evidence for an exponential cutoff
in our spectrum of 4U~0352+309.

Assuming that the bulk motion of the electrons is small, as might
be the case below an accretion shock \citep{bur91}, then the
power-law component could be a Comptonized spectrum due to the
electrons in the accretion column. In the case of 4U~0352+309 the
low accretion rate would lead to less efficient cooling of the
infalling electrons, and therefore a higher electron temperature
\citep{hay85}. In this model the exponential cutoff is at the
electron temperature \citep{ryb79}, which could explain the high
energy ($\gtrsim60$\,keV) of any roll over in the spectrum: the
temperature of the infalling electrons has increased the cutoff
energy to beyond our sensitive detection band. Using the the
photon index \wgamma\ and cutoff energy \ecut, the electron
scattering depth $\tau_{\rm es}$ can be estimated \citep{ryb79}.
For an index of $\wgamma=1.83$ and a 60\,keV cutoff, the optical
depth to electron scattering is $\tau_{\rm es}\sim 1$.

These hotter electrons can also account for the broad width of the
\crsf. It is thought that line widths are due, at least in part,
to thermal Doppler broadening by the electrons
\citep{mes85a,mes85b}. Because the motion of the electrons
responsible for the scattering is quantized perpendicular to the
field lines but free parallel to them, the width of the line
should increase as the viewing angle approaches the field
direction. \citet{mes85a} predict that the width should go as
\[\widtheqfwhm\] where $T_{\rm e}$ is the electron temperature,
$m_{\rm e}c^{2}$ is the electron rest mass, and $\theta$ is the
viewing angle with respect to the field. If we assume that the
electron temperature is $\gtrsim60$\,keV (as indicated by the
cutoff energy), then the \crsf\ width is $\wcy{} \gtrsim
0.34\ecy{} \cos(\theta)$. \citet{del00} estimate the inclination
angle of the system to be $i\wsim23-30\degree$, so if the magnetic
and spin axes are not greatly offset then $\cos(\theta)\wsim0.9$.
Therefore the width of the \crsf, while being quite wide, is
roughly consistent with the estimates of the infalling electron
temperature and viewing geometry.

The value of $1.7\dx{30}$\,\mom\ for the dipole moment derived
from the \crsf\ is very different from what is estimated from
accretion torque theory if the pulsar is spinning in equilibrium
with the inner edge of an accretion disk \citep{gho79}. In the
case of 4U~0352+309, with 837\,s spin period and
4.2\dx{34}\,\lum\ luminosity, the implied magnetic dipole moment
is $\gtrsim$4.1\dx{31}\,\mom\ (depending on the model used for the
magnetospheric radius), or at least 24 times larger than what we
measure. Using the observed luminosity and magnetic field
strength, accretion torque theory predicts the equilibrium spin
period to be in the range 20--50\,s. However, instead of spinning
up to this period the general trend is currently towards spinning
down \citep[although it has been measured only
infrequently,][]{sal98}. One explanation for this is to assume
that the star is \emph{not} currently spinning in equilibrium
with an accretion disk. Indeed, it is unlikely that a persistent
accretion disk is even present in this system. For a disk to
form, the radius at which the captured wind material would go
into a Keplerian orbit should be larger than the magnetospheric
radius \citep{sha76,wang81,li96}. For typical wind velocities
($v\gtrsim600$\,km s$^{-1}$) and wide binary separations ($P_{\rm
orb}\gtrsim100$\,days), a disk will be unable to form. Instead
the accretion is quasi-spherical beyond the magnetospheric
radius, while inside the behavior of matter is governed by the
magnetic field. Therefore accretion torque theory does not apply
in the case of 4U~0352+309, and there is no discrepancy between
the magnetic field strength and pulse period.


\section{Summary}

We have discovered a cyclotron resonant scattering feature at
29\,keV in the spectrum of 4U~0352+309. The feature is strongest
in the peak and falling edge of the pulsed emission, but also
evident in the phase average spectrum. The \crsf\ energy implies a
magnetic field strength at the polar cap of $3.3\dx{12}$\,G.
4U~0352+309 is also the lowest luminosity pulsar in which a \crsf\
has been observed. We have fit the continuum spectrum with a
combination of a black body plus a power-law. Both of the
components exist in the on-pulse minus off-pulse spectra, showing
that they are indeed coming from the pulsar magnetic polar cap.
The inferred area of the soft component is $5.4\dx{8}$\,\cmsq,
consistent with being emission from the magnetic polar cap of the
pulsar. The power-law component, although being quantitatively
different than other, more luminous accreting \xray\ pulsars, is
at least qualitatively similar. The main differences are a
steeper photon index ($\Gamma=1.8$) and the apparent lack of an
exponential cutoff. We find that accretion torque theory does not
fit the values of luminosity, spin period, and magnetic field
strength, implying that the pulsar is not spinning in equilibrium
with an accretion disk.


\acknowledgments We would like to thank the referee for his
helpful and insightful comments. This work was supported by NASA
grant NAS5-30720 and NSF Travel Grant NSF INT-9815741.


\clearpage


\clearpage

\clearpage

\clearpage

\clearpage

\clearpage

\clearpage


\begin{table}[ht]
\caption{\label{table:obs} 4U~0352+309 Observations}
\begin{minipage}{\linewidth}
\renewcommand{\thefootnote}{\thempfootnote}
\begin{tabular}{lc|lc} \hline \hline
 Date & On-Source Time (ks) & Date & On-Source Time (ks) \\ \hline
 1998 Jul 1  & 5.00  &  1998 Sep 10 & 4.94 \\
 1998 Jul 2  & 5.16  &  1998 Sep 16 & 5.26 \\
 1998 Jul 4  & 5.16  &  1998 Sep 23 & 4.04 \\
 1998 Jul 6  & 3.36  &  1998 Sep 29 & 5.62 \\
 1998 Jul 8  & 4.60  &  1998 Oct 6  & 4.74 \\
 1998 Jul 11 & 6.42  &  1998 Oct 13 & 5.54 \\
 1998 Jul 14 & 5.78  &  1998 Oct 21 & 5.00 \\
 1998 Jul 17 & 5.06  &  1998 Oct 29 & 5.32 \\
 1998 Jul 20 & 5.46  &  1998 Nov 5  & 5.60 \\
 1998 Jul 24 & 4.62  &  1998 Nov 14 & 5.00 \\
 1998 Jul 27 & 5.68  &  1998 Nov 22 & 5.32 \\
 1998 Jul 31 & 5.42  &  1998 Nov 30 & 5.54 \\
 1998 Aug 4  & 1.24  &  1998 Dec 7  & 5.36 \\
 1998 Aug 5  & 3.30  &  1998 Dec 27 & 5.00 \\
 1998 Aug 9  & 5.00  &  1999 Jan 7  & 5.00 \\
 1998 Aug 14 & 5.00  &  1999 Jan 16 & 5.08 \\
 1998 Aug 19 & 5.48  &  1999 Jan 26 & 5.48 \\
 1998 Aug 24 & 5.36  &  1999 Feb 5  & 5.00 \\
 1998 Aug 29 & 5.00  &  1999 Feb 16 & 5.78 \\
 1998 Sep 4  & 5.00  &  1999 Feb 27 & 4.92 \\ \hline
\end{tabular}
\end{minipage}
\end{table}

\clearpage

\begin{table}[ht]
\caption{\label{table:5segfits} Five Interval Ratios}
\begin{minipage}{\linewidth}
\renewcommand{\thefootnote}{\thempfootnote}
\begin{tabular}{lccccc} \hline \hline
             & \multicolumn{5}{c}{Reduced $\chi^2$
             \footnote{See \S~\ref{sec:specvar} for explanation}} \\ \cline{2-6}
 Pulse Phase & Seg 0 & Seg 1 & Seg 2 & Seg 3 & Seg 4 \\ \hline
 Peak        & 0.784 & 1.119 & 1.446 & 0.839 & 0.923 \\
 Fall        & 1.026 & 1.212 & 0.522 & 0.663 & 0.888 \\
 Min         & 0.968 & 1.321 & 1.045 & 1.819 & 0.386 \\
 Rise        & 1.114 & 0.966 & 0.751 & 1.232 & 1.049 \\ \hline
\end{tabular}
\end{minipage}
\end{table}

\clearpage

\begin{table}
\caption{\label{table:bbodyfits} Black-Body / Power-Law fits}
\begin{minipage}{\linewidth}\small
\renewcommand{\thefootnote}{\thempfootnote}
\begin{tabular}{lccccc} \hline \hline
    & \multicolumn{4}{c}{Pulse Phase} & Phase   \\ \cline{2-5}
 Pulse Phase & Peak & Fall & Min & Rise & Average \\ \hline

 \nh (\pcmsq)\footnote{Not allowed to vary} &
 0.15  & 0.15  & 0.15  & 0.15  & 0.15  \\

 $k$T (keV) &
 1.42\wpm0.02 & $1.47_{-0.02}^{+0.03}$ & $1.43_{-0.05}^{+0.06}$ &
 $1.39_{-0.03}^{+0.04}$ & 1.45\wpm0.02 \\

 R$_{\mathrm{BB}}$ (m) &
 160\wpm30 & 130\wpm30 & 100\wpm40 & 150\wpm30 & 130\wpm30  \\

 \wgamma &
 1.79\wpm0.04 & $1.84_{-0.05}^{+0.06}$ & $1.80_{-0.08}^{+0.10}$ &
 $1.77_{-0.04}^{+0.05}$ & 1.83\wpm0.03 \\

 $F_{\wgamma}$\footnote{2-10\,keV Power-Law Flux, in units of $10^{-10}$\,\flux} &
 $1.63_{-0.06}^{+0.07}$ & $1.51_{-0.08}^{+0.13}$ & $1.1_{-0.2}^{+0.4}$ &
 $1.30_{-0.06}^{+0.08}$ & $1.35_{-0.05}^{+0.06}$ \\

 \ecy{} (keV) & $27.7_{-2.5}^{+2.1}$ & 32.7\wpm2.8 & $31\pm6$ &
 $20.9_{-1.7}^{+3.6}$ & $28.6_{-1.7}^{+1.5}$ \\

 \dcy{} & 0.64\wpm0.16 & 0.70\wpm0.20 & $0.8\pm0.4$ &
 $0.34_{-0.11}^{+0.14}$ & 0.66\wpm0.12 \\

 \wcy{} (keV) & 7.9\wpm1.5 & $11.2_{-2.5}^{+3.0}$ & $13_{-6}^{+8}$ &
 $5.7_{-1.5}^{+2.9}$ & 9.0\wpm1.3 \\

 \wchired/DOF &
 0.933/80 & 0.997/80 & 0.746/80 & 0.995/80 & 0.876/80 \\

 F-test\footnote{Improvement in fit by the addition of a \crsf} &
 $5\times10^{-15}$ & $2\times10^{-6}$ &
 $4\times10^{-5}$ & $1\times10^{-7}$ &
 $1\times10^{-18}$ \\ \hline

\end{tabular}
\end{minipage}
\end{table}

\clearpage



\begin{thebibliography}{}

\bibitem[\protect\astroncite{Barret et~al.}{2000}]{bar00}
Barret, D., Olive, J.~F., Boirin, L., Done, C., Skinner, G.~K.,
\& Grindlay,
  J.~E.,  2000, ApJ, 533, 329

\bibitem[\protect\astroncite{Becker et~al.}{1979}]{bec79}
Becker, R.~H., Boldt, E.~A., Holt, S.~S., Pravdo, S.~H., \&
Robinson-Saba, J.,
  1979, ApJ, 227, L21

\bibitem[\protect\astroncite{Bildsten et~al.}{1997}]{bil97}
Bildsten, L., et~al., 1997, ApJS, 113, 367

\bibitem[\protect\astroncite{Burnard, Arons \& Klein}{1991}]{bur91}
Burnard, D.~J., Arons, J., \& Klein, R.,  1991, ApJ, 367, 575

\bibitem[\protect\astroncite{Chakrabarty et~al.}{1997}]{cha97}
Chakrabarty, D., et~al., 1997, ApJ, 474, 414

\bibitem[\protect\astroncite{{Cusumano} et~al.}{1998}]{cus98}
{Cusumano}, G., {di Salvo}, T., {Burderi}, L., {Orlandini}, M.,
{Piraino}, S.,
  {Robba}, N., \& {Santangelo}, A.,  1998, A\&A, 338, L79

\bibitem[\protect\astroncite{Delgado-Marti et~al.}{2000}]{del00}
Delgado-Marti, H., Levine, A.~M., Pfahl, E., \& Rappaport,
S.~A.,  2000,
  astro-ph/0004258

\bibitem[\protect\astroncite{di~Salvo et~al.}{1998}]{sal98}
di~Salvo, T., Burderi, L., Robba, N.~R., \& Guainazzi, M.,  1998,
ApJ, 509, 897

\bibitem[\protect\astroncite{Frontera et~al.}{1985}]{fro85}
Frontera, F., dal Fiume, D., Dusi, W., Morelli, E., \& Spada,
G.,  1985,
  Advances~in~Space~Research, 5, 125

\bibitem[\protect\astroncite{Frontera et~al.}{1979}]{fro79}
Frontera, F., Fuligni, F., Morelli, E., \& Venturea, G.,  1979,
ApJ, 229, 291

\bibitem[\protect\astroncite{Ghosh \& Lamb}{1979}]{gho79}
Ghosh, P., \& Lamb, F.~K.,  1979, ApJ, 234, 296

\bibitem[\protect\astroncite{Haberl}{1994}]{hab94}
Haberl, F.,  1994, A\&A, 283, 175

\bibitem[\protect\astroncite{Hayakawa}{1985}]{hay85}
Hayakawa, S.,  1985, Phys. Rep., 121, 317

\bibitem[\protect\astroncite{Heindl \& Chakrabarty}{1999}]{biff99}
Heindl, W.~A., \& Chakrabarty, D.,  1999,
\newblock in Proceedings of the Symposium ``Highlights in X-ray Astronomy in
  honour of Joachim Tr\"umper's 65th birthday'', ed. B. Aschenbach, M.~J.
  Freyberg, Vol. MPE Report 272, ~25

\bibitem[\protect\astroncite{Heindl et~al.}{1999}]{hei99}
Heindl, W.~A., Coburn, W., Gruber, D.~E., Pelling, M.~R.,
Rothschild, R.~E.,
  Wilms, J., Pottschmidt, K., \& Staubert, R.,  1999, ApJ, 521, L49

\bibitem[\protect\astroncite{Jackson}{1975}]{jackson75}
Jackson, J.~D.,  1975,
\newblock Classical Electrodynamics, 2nd Edition,
\newblock  (New York: Wiley)

\bibitem[\protect\astroncite{Jahoda}{2000a}]{jah00priv}
Jahoda, K.,  2000a, private communication

\bibitem[\protect\astroncite{Jahoda}{2000b}]{jah00}
Jahoda, K.,  2000b,
\newblock in Rossi2000: Astrophysics with the Rossi X-ray Timing Explorer

\bibitem[\protect\astroncite{Jahoda et~al.}{1996}]{jah96}
Jahoda, K., Swank, J.~H., Giles, A.~B., Stark, M.~J., Strohmayer,
T., Zhang,
  W., \& Morgan, E.~H.,  1996, SPIE, 2808, 59

\bibitem[\protect\astroncite{Knight}{1982}]{kni82}
Knight, F.~K.,  1982, ApJ, 260, 538

\bibitem[\protect\astroncite{Kreykenbohm et~al.}{1999}]{kre99}
Kreykenbohm, I., Kretschmar, P., Wilms, J., Staubert, R.,
Kendziorra, E.,
  Gruber, D., Heindl, W.~A., \& Rothschild, R.,  1999, A\&A, 341, 141

\bibitem[\protect\astroncite{Li \& van~den Heuvel}{1996}]{li96}
Li, X., \& van~den Heuvel, E. P.~J.,  1996, A\&A, 314, L13

\bibitem[\protect\astroncite{{Makishima} et~al.}{1999}]{mak99}
{Makishima}, K., {Mihara}, T., {Nagase}, F., \& {Tanaka}, Y.,
1999, ApJ, 525,
  978

\bibitem[\protect\astroncite{{Makishima} et~al.}{1988}]{mak88}
{Makishima}, K., et~al., 1988, Nature, 333, 746

\bibitem[\protect\astroncite{Manchester \& Taylor}{1977}]{manchester77}
Manchester, R.~N., \& Taylor, J.~H.,  1977,
\newblock Pulsars,
\newblock  (San Francisco: W. H. Freeman)

\bibitem[\protect\astroncite{Mason et~al.}{1976}]{mas76}
Mason, K.~O., White, N.~E., Sanford, P.~W., Hawkins, F.~J.,
Drake, J.~F., \&
  York, D.~G.,  1976, MNRAS, 176, 193

\bibitem[\protect\astroncite{Mavromatakis}{1993}]{mav93}
Mavromatakis, F.,  1993, A\&A, 276, 353

\bibitem[\protect\astroncite{M\'{e}sz\'{a}ros \& Nagel}{1985a}]{mes85a}
M\'{e}sz\'{a}ros, P., \& Nagel, W.,  1985a, ApJ, 298, 147

\bibitem[\protect\astroncite{M\'{e}sz\'{a}ros \& Nagel}{1985b}]{mes85b}
M\'{e}sz\'{a}ros, P., \& Nagel, W.,  1985b, ApJ, 299, 138

\bibitem[\protect\astroncite{Mihara}{1995}]{mih95}
Mihara, T.,  1995,
\newblock {\em Ph.D. thesis\/}, University of Tokyo

\bibitem[\protect\astroncite{Murakami et~al.}{1987}]{mur87}
Murakami, T., Ikegami, T., Inour, H., \& Makishima, K.,  1987,
PASJ, 39, 253

\bibitem[\protect\astroncite{Mushotzky et~al.}{1997}]{mus77}
Mushotzky, R.~F., Roberts, D.~H., Baity, W.~A., \& Peterson,
L.~E.,  1997, ApJ,
  211, L129

\bibitem[\protect\astroncite{Nelson et~al.}{1995}]{nel95}
Nelson, R.~W., Wang, J. C.~L., Salpeter, E.~E., \& Wasserman,
I.,  1995, ApJ,
  438, L99

\bibitem[\protect\astroncite{Orlandini et~al.}{1998}]{orl98}
Orlandini, M., et~al., 1998, ApJ, 500, L163

\bibitem[\protect\astroncite{Robba et~al.}{1996}]{rob96}
Robba, N.~R., Burderi, L., Wynn, G.~A., Warwick, R.~S., \&
Murakami, T.,  1996,
  ApJ, 472, 341

\bibitem[\protect\astroncite{Robba \& Warwick}{1989}]{rob89}
Robba, N.~R., \& Warwick, R.~S.,  1989, ApJ, 346, 469

\bibitem[\protect\astroncite{Rothschild et~al.}{1998}]{rot98}
Rothschild, R.~E., et~al., 1998, ApJ, 496, 538

\bibitem[\protect\astroncite{Rybicki \& Lightman}{1979}]{ryb79}
Rybicki, G.~B., \& Lightman, A.~P.,  1979,
\newblock Radiative processes in astrophysics,
\newblock  (New York: Wiley-Interscience)

\bibitem[\protect\astroncite{{Santangelo} et~al.}{1998}]{san98}
{Santangelo}, A., {del Sordo}, S., {Segreto}, A., {dal Fiume},
D., {Orlandini},
  M., \& {Piraino}, S.,  1998, A\&A, 340, L55

\bibitem[\protect\astroncite{{Santangelo} et~al.}{1999}]{san99}
{Santangelo}, A., et~al., 1999, ApJ, 523, L85

\bibitem[\protect\astroncite{Schlegel et~al.}{1993}]{sch93}
Schlegel, E.~M., et~al., 1993, ApJ, 407, 744

\bibitem[\protect\astroncite{Shapiro \& Lightman}{1976}]{sha76}
Shapiro, S.~L., \& Lightman, A.~P.,  1976, ApJ, 204, 555

\bibitem[\protect\astroncite{Shapiro \& Teukolsky}{1983}]{shapiro83}
Shapiro, S.~L., \& Teukolsky, S.~A.,  1983,
\newblock Black holes, white dwarfs, and neutron stars: The physics of compact
  objects,
\newblock  (New York: Wiley-Interscience)

\bibitem[\protect\astroncite{Tanaka}{1986}]{tan86}
Tanaka, Y.,  1986,
\newblock in IAU Colloq. 89: Radiation Hydrodynamics in Stars and Compact
  Objects, ed. D. Mihalas, K.~H. Winkler,  (New York: Springer),  198

\bibitem[\protect\astroncite{Taylor, Manchester \& Lyne}{1993}]{tay93}
Taylor, J.~H., Manchester, R.~N., \& Lyne, A.~G.,  1993, ApJS,
88, 529

\bibitem[\protect\astroncite{Telting et~al.}{1998}]{tel98}
Telting, J.~H., Waters, L. B. F.~M., Roche, P., Boogert, A.
C.~A., Clark,
  J.~S., de~Martino, D., \& Persi, P.,  1998, MNRAS, 296, 785

\bibitem[\protect\astroncite{{Tr\"{u}mper} et~al.}{1978}]{tru78}
{Tr\"{u}mper}, J., {Pietsch}, W., {Reppin}, C., {Voges}, W.,
{Staubert}, R., \&
  {Kendziorra}, E.,  1978, ApJ, 219, L105

\bibitem[\protect\astroncite{Wang}{1981}]{wang81}
Wang, Y.,  1981, A\&A, 102, 36

\bibitem[\protect\astroncite{White, Mason \& Sanford}{1977}]{whi77}
White, N.~E., Mason, K.~O., \& Sanford, P.~W.,  1977, Nature,
267, 229

\bibitem[\protect\astroncite{White et~al.}{1976}]{whi76}
White, N.~E., Mason, K.~O., Sanford, P.~W., \& Murdin, P.,  1976,
MNRAS, 176,
  201

\bibitem[\protect\astroncite{White, Swank \& Holt}{1983}]{whi83}
White, N.~E., Swank, J.~H., \& Holt, S.~S.,  1983, ApJ, 270, 711

\bibitem[\protect\astroncite{White et~al.}{1982}]{whi82}
White, N.~E., Swank, J.~H., Holt, S.~S., \& Parmar, A.~N.,  1982,
ApJ, 263, 277

\bibitem[\protect\astroncite{{Wilms} et~al.}{1999}]{wil99}
{Wilms}, J.~., {Nowak}, M.~A., {Dove}, J.~B., {Fender}, R.~P., \&
{di Matteo},
  T.,  1999, ApJ, 522, 460

\bibitem[\protect\astroncite{Worrall et~al.}{1981}]{wor81}
Worrall, D.~M., Knight, F.~K., Nolan, P.~L., Rothschild, R.~E.,
Levine, A.~M.,
  Primini, F.~A., \& Lewin, W. H.~G.,  1981, ApJ, 247, L31

\end{thebibliography}
\end{document}